\documentclass[sigconf,screenreview,prologue,table,xcdraw,dvipsnames]{acmart}
\pdfoutput=1
\AtBeginDocument{%
  }

\setcopyright{acmlicensed}
\copyrightyear{2018}
\acmYear{2018}
\acmDOI{XXXXXXX.XXXXXXX}

\acmConference[Conference acronym 'XX]{Make sure to enter the correct
  conference title from your rights confirmation emai}{June 03--05,
  2018}{Woodstock, NY}
\acmISBN{978-1-4503-XXXX-X/18/06}

\usepackage{listings,listings-rust,listings-golang}
\definecolor{offwhite}{RGB}{255,255,242}
\usepackage{algorithm}
\usepackage{algpseudocode}
\usepackage{graphicx}
\usepackage{textcomp}
\usepackage{xcolor}
\usepackage{xspace}
\usepackage{booktabs}
\usepackage{subcaption}
\usepackage{tcolorbox}
\usepackage{makecell}
\usepackage{sansmath}
\usepackage[symbol]{footmisc}
\usepackage{url}
\usepackage{hyperref}
\usepackage{cleveref}

\newcommand{\todo}[1]{}

\newcommand{\tool}{\textsc{Fluorine}\xspace}

\newcommand{\restart}{\textbf{Restart}\xspace}
\newcommand{\hinted}{\textbf{Hinted}\xspace}

\newcommand{\repair}{\textbf{BaseRepair}\xspace}

\newcommand{\capr}{\textbf{CAPR}\xspace}
\newcommand\notsotiny{\@setfontsize\notsotiny{6}{7}}

\definecolor{prompt_bg}{RGB}{252,255,221}
\definecolor{prompt_title}{RGB}{0,51,102}

\begin{document}

\title{Towards Translating Real-World Code with LLMs:\\ A Study of Translating to Rust}

\author{Hasan Ferit Eniser*}
\affiliation{%
 \institution{MPI-SWS}
 \country{Germany}
}
\author{Hanliang Zhang*}
\affiliation{%
 \institution{University of Bristol}
 \country{UK}
}
\author{Cristina David}
\affiliation{%
 \institution{University of Bristol}
 \country{UK}
}
\author{Meng Wang}
\affiliation{%
 \institution{University of Bristol}
 \country{UK}
}
\author{Maria Christakis}
\affiliation{%
 \institution{TU Wien}
 \country{Austria}
}
\author{Brandon Paulsen}
\affiliation{%
 \institution{Amazon Web Services, Inc.}
 \country{US}
}
\author{Joey Dodds}
\affiliation{%
 \institution{Amazon Web Services, Inc.}
 \country{US}
}
\author{Daniel Kroening}
\affiliation{%
 \institution{Amazon Web Services, Inc.}
 \country{US}
}

\maketitle
\renewcommand{\shortauthors}{Eniser et al.}

Large language models (LLMs) show promise in code translation -- the task of translating code written in one programming language to another language -- due to their ability to write code in most programming languages. However, LLM's effectiveness on translating real-world code remains largely unstudied. 
In this work, we perform a comprehensive study of using LLMs to translate code to Rust, with 
an emphasis on ensuring the functional correctness of the translations. We conduct our study on code extracted from real-world GitHub projects.
To enable our study, we develop \tool{}, a framework for automated, I/O equivalent translation to Rust. 
Key to \tool{} is a cross-language differential fuzzer, which allows us to automatically obtain evidence of I/O equivalence between the input source program and the Rust translation, whereas prior work required unit tests in the target language for this purpose. 
\tool{} is instantiated with an LLM and a \textit{feedback strategy}. The LLM is first used to obtain a candidate Rust translation, which is then checked for I/O equivalence. If the original and the translated programs are not I/O equivalent, the feedback strategy is invoked, which prompts the LLM to repair the buggy translation. 
We use \tool{} to assess the ability of five state-of-the-art LLMs (GPT4, Claude~3, Claude 2.1, Gemini Pro, and Mixtral) to translate code to Rust, and we experiment with four different feedback strategies to assess the LLMs' ability to repair buggy translations.
Our results show that the most successful LLM can translate 47\% of our benchmarks, and also provide insights into next steps for improvements. Our artifact is available at \url{https://d34gtk3knhjgeg.cloudfront.net/artifact.tar.gz}.
\section{Introduction}
\label{section:intro}

The task of translating programs between languages is growing increasingly important due to the rising interest in safe programming languages like Rust and the need to modernize potentially buggy legacy code by converting it into such languages~\cite{tractor}. This effort is particularly relevant for AWS, which has legacy code requiring such updates.
Currently, this translation is performed manually, but the process is both time-consuming and error-prone.

Recently, LLMs have been explored for this purpose~\cite{tang-etal-2023-explain,RoziereLachaux2020,RoziereZhang2022,szafraniec2022code,DBLP:journals/corr/abs-2407-07472}. However, these efforts have primarily focused on code sourced from competitive programming websites~\cite{puri2021codenet}, educational websites~\cite{ahmad2021avatar}, or hand-crafted coding problems~\cite{liu2024your,chen2021evaluating}. Few attempts have been made to translate real-world code. Out of these, Shiraishi et al.~\cite{DBLP:journals/corr/abs-2409-10506} translate C to Rust, but focus exclusively on generating compilable code. Ibrahimzada et al.~\cite{DBLP:journals/corr/abs-2410-24117} translate Java to Python, and validate I/O equivalence for only 25.8\% of the translated functions. Lastly, Zhang et al. 
focused on small code changes~\cite{zhang2023multilingual}.
Hand-crafted benchmarks are not representative of real-world code. They typically comprise of a single function using only primitive data types, whereas real-world code features many functions and complex, user-defined data types.

In this work, we take a further step toward answering the question whether LLMs can translate real-world code.
To answer this question,
we design and implement a framework for automatically producing validated Rust translations~\cite{DBLP:conf/tacas/PnueliSS98} with an LLM, which we use to conduct 
a comprehensive study on leveraging LLMs for this task. 
Compared to translating smaller, hand-crafted benchmarks, there are two key challenges that need to be addressed when targeting real-world code.


\paragraph{Challenge 1: Real-world project size} Generally, entire real-world projects cannot fit in the context window of the LLMs.
To address this challenge, we extract code samples from real-world projects that can fit in the context window of LLMs. Our code samples contain between 1 and 25 functions, and up to 598 lines of code. To keep our scope practical, our code samples only use standard libraries. These code samples are still illustrative of real-world code as they contain features such as global variables, user-defined, dynamically-allocated, data structures, array pointers, type casts, enumeration types, error handling etc. (see the examples later in the section). To automate and reduce bias in the selection of code samples, we develop a methodology and tool for extracting them from projects. 

\paragraph{Challenge 2: Lack of unit tests in the target language}
For a translation to be useful, it must be semantically equivalent to the original code. 
In prior work, semantic equivalence is typically verified by running the translated program against existing unit tests in the target language,
but in practical settings, unit tests in the target language would not already exist.
One solution would be to either hand-write unit tests in the target language, but this is a tedious and time-consuming process. 
Another solution is to automatically translate unit tests from the source project, but this is non-trivial as it requires mapping data between different languages.
Moreover, even if unit tests from the source language can be translated, they may lack sufficient coverage to reliably validate correctness of the translation.
To overcome the lack of unit tests, we develop a cross-language fuzzer capable of automatically passing inputs and outputs between languages.

\paragraph{Our framework for automated, validated Rust translation}
We use our cross-language fuzzer to build \tool{}, a framework for automatically producing (validated) semantically equivalent Rust translations. Our cross language differential fuzzer allows us to obtain evidence of semantic equivalence without unit tests. 
\tool{} is instantiated with two key components: an LLM and a \textit{feedback strategy}.
The LLM is first prompted to write a candidate translation of the source program, and if the translation does not compile, it is prompted to repair compilation errors. Once the translation compiles, we invoke our cross-language differential fuzzer, which tests the translation for equivalence to the original, and returns counterexamples if the translation is not equivalent. If a counterexample is found, the \textit{feedback strategy} is invoked. The feedback strategy takes the counterexample as input, and returns a prompt for the LLM to repair the buggy translation. Once the fuzzer fails to find a counterexample after a configured timeout, the translation is considered validated.

\paragraph{Benchmarks.} 
For the benchmarks, we extract code samples from seven open source projects written in C and Go. We choose these languages because, at AWS, we have many C and Go codebases, and rewriting them in Rust offers certain advantages. Compared to C, Rust offers memory safety. Compared to Go, Rust uses much less memory as it does not include a runtime in its executable. The open source projects are from a diverse set of domains: audio processing, text processing, geometry, banking, 2D triangulation, graph algorithms, and sound card emulation. 

The extracted code samples include complex features that are typically absent in hand-crafted datasets.
For example, Figure \ref{fig:tnf-go} contains a program extracted from the \emph{ACH} library featuring 
a global variable \lstinline[language=go]{moov_io_ach_stringZeros}, which is initialised with the function call 
\lstinline[language=go]{moov_io_ach_populateMap(94, "0")}. This kind of initialization of a global variable is not allowed in Rust, making it non-trivial to find an equivalent translation without resorting to unsafe code.
Claude~3 managed to find the following translation:
\begin{lstlisting}[language=Rust, basicstyle=\scriptsize\ttfamily]
static MOOV_IO_ACH_STRING_ZEROS: 
    Lazy<HashMap<usize, String>> = 
    Lazy::new(|| populate_map(94, "0"));
\end{lstlisting}
This snippet uses \lstinline[language=Rust]{once_cell::Lazy}, which stores a value that gets initialized on the first access.

As another example, Figure \ref{fig:evnadd-go} contains a program we extracted from the \emph{go-gt} library, featuring a user-defined type \lstinline[language=go]{Env} that assembles several arrays, pointers and numeric data. Mapping \lstinline[language=go]{Env} to an exact counterpart in Rust is not obvious, as  
a slice \lstinline[language=go]{[]int64} in Golang can either be represented by a vector in Rust \lstinline[language=Rust]{Vec<i64>}, which is a growable, owning array-like data type, or a borrowed-slice \lstinline[language=Rust]{&'a [i64]}, a non-growable, non-owning array-like data type.
Our cross-language differential fuzzer
handles translations of the function \lstinline[language=go]{add} that use both variants for \lstinline[language=go]{Env} by correctly mapping between the Go and Rust representations of its inputs (where the receiver \lstinline[language=go]{e} of type \lstinline[language=go]{*Env} is one of the inputs).  

The code in Figure \ref{fig:lcs-go}, extracted from the \emph{go-edlib} library, returns all the longest common subsequences of two input strings (for brevity, we omit the callees). This code presents several challenges, for instance, finding a correct mapping between different styles of error handling. In Golang, a failable computation output is typically expressed by a pair of the target output type and the error type, as shown in the signature of \lstinline[language=go]{LCSBacktrackAll}. 
On the other hand, in Rust, this is often expressed by an optional output type, such as \lstinline[language=Rust]!Result<Vec<String>, Error>!. Moreover, this program contains casts from strings to arrays, and array manipulation, which need to be correctly mapped to their corresponding Rust representation. 

\begin{figure}
\begin{lstlisting}[language=Go, basicstyle=\scriptsize\ttfamily]
var (
	moov_io_ach_stringZeros map[int]string = moov_io_ach_populateMap(94, "0")
)

func moov_io_ach_populateMap(max int, zero string) map[int]string {
	out := make(map[int]string, max)
	for i := 0; i < max; i++ {
		out[i] = strings.Repeat(zero, i)
	}
	return out
}
\end{lstlisting}
\vspace{-0.3cm}
\caption{Code sample from \emph{ACH}}
\label{fig:tnf-go}
\vspace{-0.5cm}
\end{figure}


\begin{figure}
\begin{lstlisting}[language=Go, basicstyle=\scriptsize\ttfamily]
func (e *Env) add(i, p int64) {
	var j int64
	e.S[i] = true
	e.Prev[i] = p
	for j = 0; j < e.N; j++ {
		if e.Lx[i]+e.Ly[i]-e.G.Get(i, j) < e.Slack[i] {
			e.Slack[i] = e.Lx[i] + e.Ly[i] - e.G.Get(i, j)
			e.Slackx[i] = j
		}
	}
}

func (m Matrix) Get(i int64, j int64) int64 {
	return m.A[i*m.N+j]
}

type Env struct {
	N                           int64
	G                           *Matrix
	S                           []bool
	Slack, Slackx, Prev, Lx, Ly []int64
}

type Matrix struct {
	N int64
	A []int64
}
\end{lstlisting}
\vspace{-0.3cm}
\caption{Function \texttt{add} from \emph{go-gt}}
\label{fig:evnadd-go}
\vspace{-0.2cm}
\end{figure}

\begin{figure}
\begin{lstlisting}[language=Go, basicstyle=\scriptsize\ttfamily]
func LCSBacktrackAll(str1, str2 string) ([]string, error) {
	runeStr1 := []rune(str1)
	runeStr2 := []rune(str2)

	if len(runeStr1) == 0 || len(runeStr2) == 0 {
		return nil, errors.New("Can't process and backtrack any LCS with empty string")
	} else if Equal(runeStr1, runeStr2) {
		return []string{str1}, nil
	}
	return processLCSBacktrackAll(
		str1,
		str2,
		lcsProcess(runeStr1, runeStr2),
		len(runeStr1),
		len(runeStr2),
	).ToArray(), nil
}
\end{lstlisting}
\vspace{-0.3cm}
\caption{Function \lstinline{LCSBacktrackAll} from \emph{go-edlib}}
\label{fig:lcs-go}
\vspace{-0.2cm}
\end{figure}

\begin{figure}
\begin{lstlisting}[language=Rust, basicstyle=\scriptsize\ttfamily]
struct Env {
    n: i64,
    g: Box<Matrix>,
    s: Vec<bool>,
    slack: Vec<i64>,
    slackx: Vec<i64>,
    prev: Vec<i64>,
    lx: Vec<i64>,
    ly: Vec<i64>,
}

struct Matrix {
    n: i64,
    a: Vec<i64>,
}

fn add(e: &mut Env, i: i64, p: i64) {
    let mut j: i64 = 0;
    e.s[i as usize] = true;
    e.prev[i as usize] = p;
    for j in 0..e.n {
        if e.lx[i as usize] + e.ly[i as usize] - get(&e.g, i, j) < e.slack[i as usize] {
            e.slack[i as usize] = e.lx[i as usize] + e.ly[i as usize] - get(&e.g, i, j);
            e.slackx[i as usize] = j;
        }
    }
}

fn get(m: &Matrix, i: i64, j: i64) -> i64 {
    m.a[(i * m.n + j) as usize]
}
\end{lstlisting}
\vspace{-0.3cm}
\caption{Rust translation of function \lstinline{add} from \emph{go-gt}}
\label{fig:evnadd-rust}
\vspace{-0.5cm}
\end{figure}


\paragraph{Evaluation.}
\todo{update experiment numbers}
We instantiate our framework with five state of the art LLMs---GPT-4o, Claude 3, Claude 2.1, Gemini Pro, and Mixtral---and four different feedback strategies. Three of our feedback strategies leverage the counterexamples to construct different prompts to repair the buggy translation. The fourth strategy is a  baseline that simply repeats the original prompt, hoping randomness in LLM inference would yield a different candidate translation. We execute each LLM and feedback strategy on 408 extracted code samples, for a total of 8160 code translation experiments.

\paragraph{Key findings.}
Our experiments led to several key findings, which we discuss in detail in the rest of the paper:

(1) The tested LLMs are able to translate a substantial number of our code samples, achieving  overall success rates of 47.3\% (GPT-4o), 44.2\% (Claude 2), 38.5\% (Claude 3), 33.8\% (Gemini Pro), and 19.5\% (Mixtral) in terms of percentage of benchmarks correctly translated. As a trend, increasing complexity, especially in terms of lines of code, reduces success rates. In particular, success rates tend to drop off somewhere around 100 lines of code. 

While fully translating entire projects is an ongoing challenge, these results are promising: they show LLMs have the ability to translate portions of real-world projects, which can significantly reduce the manual effort and cost associated with such tasks. At AWS, this could potentially reduce the manual effort and costs involved in such tasks. 

(2) LLMs generally produced high quality Rust code, as reported by Clippy~\cite{clippy}, Rust's standard linter. However, occasionally (1--15\% of the time), the translated code could be more idiomatic (Clippy reported a few style warnings), more concise (complexity warnings), or more performant (performance warnings). This is particularly important for a potential AWS use case, where code quality and maintainability are critical concerns.

(3) The most effective feedback strategy is to re-generate a new translation from scratch, yielding an improvement of 7--21\% on our benchmarks. This simple strategy is more effective than more complex strategies that involve prompting the LLM to repair a translation with counterexamples. This suggests that LLMs are not always able to understand counterexamples. We discuss this trend further in Section~\ref{sec:feedback-disucssion}.

\noindent\emph{Contributions.} We make the following contributions:
\begin{itemize}
    \item We develop \tool{}, a framework for producing validated Rust translations without the need for hand-written test cases 
    \item We use \tool{} to conduct the first substantial study of using LLMs to translate real-world code
    \item We demonstrate that LLMs are capable of translating parts of real-world projects, and that directly providing counterexamples as feedback to an LLM is less effective than repeating the original prompt
\end{itemize}

\section{Related Work}
\label{section:related_work}
In this section, we discuss closely related work from the literature under several categories. 

\textbf{Code Translation.}
The most closely related code translation works use LLMs for translation where the source and target languages are different.
Most of them~\cite{tang-etal-2023-explain, RoziereLachaux2020, RoziereZhang2022, szafraniec2022code,jana2023attention} evaluate exclusively on competitive programming style code. 
Some works have explored real-world code. For instance, Zhang et al.~\cite{zhang2023multilingual} evaluate on real-world API code, translating Java to C\#, though their approach requires additional fine-tuning, unlike ours. Pan et al.~\cite{PanICSE24} fail to produce syntactically correct code for such examples. Shiraishi et al.~\cite{DBLP:journals/corr/abs-2409-10506} focus exclusively on generating compilable code rather than ensuring functional correctness. Ibrahimzada et al.~\cite{DBLP:journals/corr/abs-2410-24117} translate Java to Python, validating I/O equivalence for only 25.8\% of the translated functions. 
Two of the works~\cite{PanICSE24,jana2023attention} conclude that counterexamples can be useful feedback, which does not match our conclusion. We compare our results with theirs in Section~\ref{sec:feedback-disucssion}. 

Other LLM/ML code translation works focus on problems where the source and target language are the same~\cite{zhang2022heterogen,mariano2022automated}. We consider this a different task than ours, because the goals are different. Meta studies have been conducted on code translation as well in ~\cite{eniser2023automatically,jiao2023evaluation}, though they do not provide insight on translating real-world code.
Finally, several works have developed rule-based techniques for specific source and target language pairs such as C to Rust,~\cite{zhang2023ownership, emre2021translating,c2rust}, C to Go~\cite{cgo}, and Java to C\#~\cite{sharpen}. While rule-based approaches can theoretically guarantee correctness of the translation, they require significant engineering effort to build, and they can produce unidiomatic code as we demonstrate in our results.

\textbf{Cross-Language Differential Fuzzing.}
While differential fuzzing and testing has a rich literature, the majority do not consider comparing implementations in two different languages. There are many works that compare programs in the same language using symbolic execution~\cite{noller2020hydiff, bohme2013regression, palikareva2016shadow, person2011directed} and fuzzing~\cite{guo2018dlfuzz, jin2010automated, nilizadeh2019diffuzz, petsios2017nezha}. Such works do not need to solve the problem of mapping data from one language to another, though they are likely complementary to our work---they could be used to improve the coverage achieved by our fuzzer. Works in fuzzing multi-language systems~\cite{li2023polyfuzz} do not address this problem either. Only one work~\cite{garzella2020xlverify} attempts general cross-language testing like we do by compiling both languages down to a shared IR. As we will discuss in Section~\ref{section:oracle}, this approach cannot effectively handle user-defined data types, and is heavily dependent on the IR compiler preserving structure of the original source program. 

\textbf{Feedback Strategies for LLMs.} 
Relatively few works have focused on developing feedback strategies for LLMs.
Among them, recent work in automated program repair~\cite{xia2023automated,kong2024contrastrepair} reports success with an approach that provides counterexamples as feedback. We discuss their results in relation to ours in Section~\ref{sec:feedback-disucssion}. Another recent approach proposes giving the LLM the entire faulty execution trace rather than just the inputs~\cite{DBLP:conf/icml/NiACDSSY24}. However, this method requires fine-tuning the LLM to reason effectively with execution traces, which we are unable to do with the models tested in this work. Instead of using feedback strategies to re-query the LLM for a refined translation, Yin et al.~train a second LLM to fix translation errors~\cite{DBLP:journals/corr/abs-2407-07472}, a direction we may explore as future work. While any automated program repair technique could be used as a feedback strategy, we focus only on feedback strategies that use an LLM to fix errors.



 

\section{Overview}
\label{section:overview}

In this section, we define the task of code translation, provide an overview of our framework for validated code translation with LLMs, and then illustrate on a concrete example.

\subsection{Code Translation}
We first formally define the problem of code translation. Let $l$ be a programming language, and $P_l$ the set of all valid programs written in $l$. Assume we have a program $p \in P_l$ that we wish to translate to a different language $l'$. That is, we wish to find $p' \in P_{l'}$ that has the same behavior as $p$ with respect to a mapping between the values of $l$ and $l'$. 

In our work, a program (i.e., $p$ or $p'$) is a set of functions, user-defined types (e.g., struct definitions), global variables, import/include statements etc. One of the functions in a program is the \textit{entry point} function. Note that the entry point function is \textit{not} necessarily \lstinline{main()}---the inputs and outputs of the entry point could be primitive data types, user-defined types (e.g. structs, classes), and even pointer types.

For simplicity of notation, we define $p$ and $p'$ as operating on \textit{program states}. A program state contains the values of the inputs outputs of the program, as well as variables defined in the global scope. Letting $S_p$ and $S_{p'}$ be the set of all program states for $p$ and $p'$, respectively, we have $p : S_p \to S_p$ and $p' : S_{p'} \to S_{p'}$. We write $p(s_{in}) = s_{out}$ where $s_{in}, s_{out} \in S_p$ to denote the result of executing $p$ on $s_{in}$.

To complete our definition of code translation, we define $M : S_p \to S_{p'} $ and $ M' : S_{p'} \to S_p$, which are mapping functions that map states of program $p$ to states of $p'$, and vice versa. Formally, translation's goal is to discover a program $p'$ such that:
\[\forall s \in S_p . p(s) = M'(p'(M(s)))\]




\subsection{Our Framework for Automated and Validated Code Translation}
Next, we describe our framework for  validated translations, \tool{}. Again, we 
assume we have a source program $p$, and we wish to discover a translation $p'$ with the same behavior.

The framework needs to be instantiated with the following components:
\begin{itemize}
\item An LLM, $G : Q \to P_{l'} $, that takes a natural language query $q \in Q$ and outputs a candidate translation $p' \in P_{l'}$. $q$ contains the original source program $p$ and natural language instructions to translate $p$ into the target language $l'$. 
\item A feedback strategy $\textsc{feedback}(q, p', E^-, E^+)$ which takes the query $q$, the candidate translation $p'$, as well as positive and negative examples, $E^+$ and $E^-$, and returns a new query that can be provided to $G$ to generate a new candidate translation. 
\end{itemize}

Note that in practice, the resulting $p'$ may have a top-level function whose function signature is incompatible with $p$ and therefore the mapping functions $M, M'$ cannot be defined,
or the program output by the LLM may not compile. We find that the former rarely happens, and we address the latter through a compilation repair phase, which is based on the approach in~\cite{deligiannis2023fixing}.

The framework contains a
\textit{cross-language differential
fuzzer}. We define this as $\textsc{fuzzer}(p, p')$, which takes the original source program and a translation, and returns two sets of examples $E^+$ and $E^-$. $E^+$ is a set of positive examples where $p$ and $p'$ agree. Positive examples have the form $(s_{in}, s_{out})$, where $s_{in}, s_{out} \in S_{p'}$. $E^-$ is the set of counterexamples where the output produced by $p$ disagrees with $p'$. A counterexample is a triple of states from $S_p'$ of the form $(s_{in}, s_{exp}, s_{act})$, which are the initial state, expected output state, and the actual output state.



The routine for validated code translation with \tool{} is shown in~\ref{alg:genworacle}. We first use $G$ to generate a candidate program $p'$ from the initial query $q$, which we then pass to the compilation driven repair routine. If this is unsuccessful at making $p'$ compile, we exit the loop and fail. Otherwise, we invoke the fuzzer to check for counterexamples. If none are found, we assume $p'$ is correct, and return it. Otherwise, we invoke a feedback routine, which generates a new $q$, and repeat the process until a program is found that passes the fuzzer check, or until some fixed budget is reached and we fail.

\begin{algorithm}
\caption{Algorithm for Validated Code Translation with \tool{}}
\label{alg:genworacle}
\begin{algorithmic}[1]
  \Require $p$ : The program to translate, $q$: The initial translation prompt, $G$: An LLM, \textsc{feedback}: A feedback strategy, $b$: A budget
  \While{$b > 0$}
    \State $p' \gets G(q)$
    \If{$\neg\textsc{compilation-repair}(p')$}
    \State \textbf{break}
    \EndIf
    \State $E^-, E^+ \gets \textsc{fuzzer}(p, p')$
    \If{$E^- = \emptyset $}
      \State \textbf{return} $p'$
    \EndIf
    \State $q \gets \textsc{feedback}(q, p', E^-, E^+)$
    \State $b \gets b - 1$
\EndWhile
\State \textbf{return} $\textsc{FAIL}$
\end{algorithmic}
\end{algorithm}

\subsection{Motivating Example}


We now illustrate our Rust translation approach with the 
concrete example in Figure~\ref{fig:evnadd-go}.
In the example, our source program $p$ is the function \lstinline{add} from the \emph{go-gt} library.
This is a subroutine of the Hungarian algorithm \cite{DBLP:books/daglib/p/Kuhn10} for finding maximum matching, which adds two edges to an alternating path during the search, and records the output by mutating the receiver \lstinline{e}.

We first create an initial query containing the Go code and instructions describing the translation task, which is given to the LLM to generate a candidate translation. If we continue past compilation driven repair, the candidate translation $p'$ in Figure~\ref{fig:evnadd-rust} is guaranteed to compile, but not to be I/O equivalent to the original source program. 
To check for I/O equivalence, $p$ and $p'$ are passed to the cross-language differential fuzzer, which generates input states, executes both programs, and checks that they produce the same output state. We capture side-effects by comparing whole program states, rather than just the explicit output.

One of the challenges that we face is executing $p$ and $p'$ in two different languages on matching input states, and then comparing their output state. Specifically for our running example, we must convert primitive types as well as user-defined data structures: \lstinline{Env} has distinct representations in Go and Rust; arguments \lstinline{i} and \lstinline{p} have type \lstinline[language=Go]{int64}
in Go, but \lstinline[language=Rust]{i64} in Rust;  \lstinline{e} is a pointer to an \lstinline{Env} value in Go, but a mutable reference in Rust.

To solve this challenge, we develop a technique based on serializing then de-serializing to exchange data between languages. We use the JSON~\cite{json} format, because most languages support it. Most data types, including complex data types and pointers can be automatically serialized into JSON, thus it allows us to easily support real-world code.
For our example, we generate inputs for the Rust candidate 
in Figure~\ref{fig:evnadd-rust}, which are then serialized to JSON and deserialized to Go. Fig.~\ref{fig:input-add-json-input} denotes such a serialized input state. Once the original Go program and its Rust translation are executed, the Go output state is
again serialized to JSON, deserialized to Rust, and compared against the Rust output state. For our example, the output state obtained by executing the Go code is given in Fig.~\ref{fig:input-add-json-output} (same as the input state in Fig.~\ref{fig:input-add-json-input}, with the only difference that the last element of field \lstinline{s} is
set to \lstinline{true} instead of \lstinline{false}). After running the cross-language differential fuzzer, we conclude that the translation in Figure~\ref{fig:evnadd-rust} computes the expected output state for all the generated input states, and it is thus deemed I/O equivalent to the original Go code, and returned by \tool{}.

\lstdefinestyle{json-pretty}{
    basicstyle=\footnotesize\ttfamily,    
    string=[s]{"}{"},
    stringstyle=\color{blue},
    comment=[l]{:},
    commentstyle=\color{black},
}
\begin{figure}
\centering
\begin{subfigure}{.45\textwidth}
\begin{lstlisting}[style=json-pretty]
{"e": {
    "n": 3,
    "g": {
      "n": 3,
      "a": [0, 0, 0, 0, 0, 0, 0, 0, 0]
    },
    "s": [false, false, false],
    "slack": [0, 0, 0],
    "slackx": [0, 0, 0],
    "prev": [0, 0, 0],
    "lx": [0, 0, 0],
    "ly": [0, 0, 0]
  },
  "i": 2,
  "p": 0}
\end{lstlisting}
\vspace{-0.2cm}
\caption{Serialized JSON input state for function \lstinline{add} from \emph{go-gt}}
\label{fig:input-add-json-input}
\end{subfigure}\qquad
\begin{subfigure}{.45\textwidth}
\begin{lstlisting}[style=json-pretty]
{"e": {
    "n": 3,
    "g": {
      "n": 3,
      "a": [0, 0, 0, 0, 0, 0, 0, 0, 0]
    },
    "s": [false, false, true],
    "slack": [0, 0, 0],
    "slackx": [0, 0, 0],
    "prev": [0, 0, 0],
    "lx": [0, 0, 0],
    "ly": [0, 0, 0]
  },
  "i": 2,
  "p": 0}
\end{lstlisting}
\vspace{-0.2cm}
\caption{Serialized JSON output state for function \lstinline{add} from \emph{go-gt}}
\label{fig:input-add-json-output}
\end{subfigure}
\end{figure}

Conversely, if a counterexample is discovered by the fuzzer, then we invoke a feedback method, which uses the counterexample to create a new query to the LLM and generates a new candidate translation. Designing a suitable feedback method is another challenging aspect of the translation task. There are many ways to re-query the LLM for a new translation, each with their own likelihood of success. Moreover, most state-of-the-art LLMs are operated as API services which charge per input token, so different query strategies will have different dollar costs. To address this, we propose and evaluate a set of feedback strategies as described in Section~\ref{section:fallback}.

\section{LLM-Based Code Translation}
\label{section:code-translation}




\subsection{Obtaining Translations}
\label{section:obtain-translation}

As mentioned in the previous sections, we are considering the problem of translating a program written in C or  Go to Rust. We use zero-shot prompting and follow the best practices given by the LLM's provider. We construct the initial query $q$ (to be input to the LLM) as sketched in \Cref{figure:translate-prompt}.

We start with a preamble describing the overall task. Then, we supply the program to be translated, and, finally, we provide specific constraints to be followed by the LLM. In particular, we have three types of constraints: formatting guidelines, code characteristics and fuzzer constraints. Formatting guidelines describe how the generated code should look, simplifying parsing and extraction of relevant information from the response. For code characteristics, we instruct the LLM to produce safe Rust code, and to maintain the same function names, parameter names, and return types from the input code. Finally, the fuzzer constraints ensure that the generated code can be handled by our fuzzer (more details on this in \Cref{section:oracle}).

The translation generated by the LLM may not initially compile. We address this with approach in~\cite{deligiannis2023fixing}. At a high level, we iteratively query the LLM to fix the error, until the code becomes compilable. Each time, we provide both the faulty translation and the error message from the Rust compiler to the LLM, and ask it to use a specific format for the suggested fixes, applying them only to the affected lines of code.

\begin{figure}
    \centering    
    \begin{tcolorbox}[
        colback=prompt_bg,
        colframe=prompt_title,
        subtitle style={boxrule=0.4pt, colback=yellow!50!blue!25!white, colupper=black},
    ]
    \footnotesize
    \textbf{Human:}\\
    \textit{\# Preamble}\\
    \texttt{You are given a C/Go program. We need to translate it to Rust.}\\
    
    \textit{\# Code to be translated}\\
    \texttt{\{C/Go Program\}} \\
    
    \textit{\# Instruction}\\
    \texttt{Give me a Rust translation of the above C/Go code.}
    
    \textit{\# Constraints}\\
    \texttt{Here are some constraints that you should respect:}
    \begin{itemize}
        \item \texttt{Give me only the translated code, don't add explanations or anything else.}  \textit{\# formatting guideline}
        \item \texttt{Use only safe Rust.}\qquad\qquad\qquad\quad\textit{\# code characteristic}
        \item \texttt{Do not use custom generics.} \qquad\textit{\# fuzzer limitation}
        \item \texttt{\dots} \\
    \end{itemize}

    \textbf{Assistant:}
    \end{tcolorbox}
    \vspace{-0.5cm}
    \caption{LLM Prompt for obtaining translations.}
    \label{figure:translate-prompt}
\end{figure}


\vspace{-0.2cm}
\subsection{Checking Translations}
\label{section:oracle}

To test the I/O equivalence between the original source program $p$ and a candidate Rust translation~$p'$, we develop a cross-language differential fuzzer. 
For a given $p$ and $p'$, we automatically generate a fuzzing harness in Rust, which uses  Bolero and libfuzzer~\cite{serebryany2016libfuzzer} to perform fuzzing. The test harness generates program states from $S_{p'}$, which are directly invoked on $p'$. We implement the mapping function $ M' : S_{p'} \to S_p $, using JSON de/serialization. We serialize the Rust program state $s'$ into JSON format, and then instrument the source program $p$ to deserialize the JSON into a program state of $S_p$. The instrumented $p$ is invoked on the serialized $s'$ from Rust using a foreign function interface. To compare outputs, we map the output state of $p$ to one of $p'$ using JSON de/serialization as well, which can then be directly compared. 

We use JSON serializers for two reasons. First, the mapping between fields of user-defined data types in the source and target language are automatically determined based on the field names. This requires the LLM to produce data types with field names that match the source program, but in our benchmarks LLMs always do this. Second, most languages support automatic serialization of primitive, pointer, and user-defined types.

We note an alternative approach, taken by~\cite{garzella2020xlverify}, is to compile both $p$ and $p'$ down to a common IR, such as LLVM, and then perform fuzzing on the IR. However, we find that IR compilers for different languages typically discard type and layout information (e.g. user-defined data types are represented as a void pointer). This makes it nearly impossible for a fuzzer to generate any meaningful inputs.

\textbf{Soundness \& Limitations}. 
Our fuzzer can only make heuristic based guarantees (e.g. coverage) on the equivalence of $p$ and $p'$. This is a limitation of fuzzing and testing in general. However, our fuzzer achieves an average line coverage of 97\%.


In addition, JSON serialization is not automatically supported for all types. For example, features in Rust like trait definitions, \textsc{impl} traits, and lifetimes in data type definitions are only partially supported. This means that the equivalence check may fail because serialization fails. We report these errors in Section~\ref{sec:results}. In addition, we do not support features like concurrency, network, and file I/O. Our benchmark selection excludes these features.


\vspace{-0.2cm}
\section{Feedback Strategies}
\label{section:fallback}
In this section, we present four feedback methods that can be used if the fuzzer finds a counterexample $E^{-}$ for the correctness of the translation $p'$ by the LLM in Alg. \ref{alg:genworacle}.

\textit{Simple Restart (\restart)} We discard the generated code $p'$ and re-query the model with the same prompt $q$. 

\textit{Hinted Restart (\hinted)} This builds on the previous strategy by adding positive and negative examples from the fuzzer, $E^+$ and $E^-$, to the original prompt $q$. 
The intention is to suggest desirable behaviours to the LLM,  as well as known faulty cases to avoid. 
We separately group the examples in $E^{+}$ and $E^{-}$ based on the paths they exercise in $p'$. Intuitively, this corresponds to splitting them into equivalence classes, where each equivalence class corresponds to a particular program path. Then, the query constructed by \hinted only contains positive and negative examples from a single equivalence class, respectively. 

\textit{Counterexample-Guided Repair (\repair)}
Discarding the generated code $p'$ when the fuzzer check fails may not always be the optimal choice. For instance, if $p'$ is close to passing the fuzzer, trying to repair it might work better. 
As part of \repair, we give counterexamples from the fuzzer to the LLM.  
%
%
Similarly to \hinted, a query only contains negative examples from the same equivalence class, which correspond to bugs associated with the same program path. The expectation is that the candidate translation generated in the next iteration of Alg. \ref{alg:genworacle} will produce the correct outputs for the given examples. A sketch of the prompt used for \repair is given in \Cref{fig:fix-prompt} (excluding the lines colored in 
{\color{magenta}magenta}).  
In Alg. \ref{alg:genworacle}, if the translation generated by $G$ for the query $q$ constructed by \repair still fails the fuzzer check, then this last faulty translation will be considered by the next call to \repair.


\begin{figure}
    \centering    
    \begin{tcolorbox}[
        colback=prompt_bg,
        colframe=prompt_title,
        subtitle style={boxrule=0.4pt, colback=yellow!50!blue!25!white, colupper=black},
    ]
    \footnotesize
    \textbf{Human:}\\

    \textit{\# Preamble}\\
    \texttt{You are given a C/Go program and its faulty Rust translation. We need to repair the faulty Rust program.}\\

    \textit{\# Code to be translated}\\
    \texttt{\{C/Go Program\}} \\
    
    \textit{\# Code to be repaired}\\
    \texttt{\{Faulty Rust Program\}} \\
    
    \textit{\# Instruction}\\
    \texttt{Make changes to the given code to obtain expected outputs for the given test inputs.}\\

    \textit{\# Constraints}\\
    \texttt{Here are some constraints that you should respect: \dots}\\

    \textit{\# Counterexamples}\\
    \texttt{CE1}\\ \texttt{CE2}\\
    
    
    \textbf{Assistant:}\\
    {\color{Magenta}
    \texttt{\{LLM generated rust translation\}}\\

    \textbf{Human:}\\
    \texttt{That is incorrect on the following inputs:}\\
    \texttt{\# Counterexamples\\CE1\\ CE2\\
    }

    \textbf{Assistant:}\\
    
    }
    \end{tcolorbox}
\vspace{-0.5cm}
    \caption{LLM Prompt for \repair and \capr. \repair is shown in black. \capr is shown in black and {\color{magenta}magenta}.}
    \label{fig:fix-prompt}
\end{figure}

\textit{Conversational Repair (\capr)}
Recent work in code translation~\cite{PanICSE24} and automated program repair~\cite{xia2023conversational}, have proposed \textit{conversational} repair approaches, wherein previous incorrect code is included in the prompt to the LLM to discourage the LLM from producing the same code again. The \capr approach begins with the same prompt as \repair, however they differ if the new translation still fails the fuzzer check. In \repair, we create a new prompt from scratch, but in \capr, we keep the prompt, and append a new piece of dialogue to it as shown in {\color{magenta} magenta} Figure~\ref{fig:fix-prompt}. This process can be repeated multiple times, meaning the prompt is a dialogue of failed translations.

The methods \restart and \hinted cost less than \repair and \capr as they don't include the incorrect translation in the prompt. Therefore the former use about half the input tokens of the latter.


\section{Evaluation}

\label{section:evaluation}

In this section, we present our results for the following research questions.

\textit{\textbf{RQ1: How do LLMs perform on translating code taken from real-world projects to Rust?}}
We gather an extensive set of benchmarks by extracting representative code samples from real-world projects, covering nearly all data structures (type definitions) and global variable usages within these projects. We use LLMs to generate translations which are then checked for correctness by the fuzzer, and fixed if needed by applying feedback strategies. We answer the following concrete questions.  

(\textit{RQ1.1}) How many benchmarks can each LLM translate from each of our projects? We report the percentage of benchmarks from each project that are successfully translated for each LLM. We show that success rates vary widely based on the benchmark and LLM. LLMs achieve up to 80\% success rate on benchmarks from our ``easiest'' project, and between 15--40\% on our ``hardest'' project.

(\textit{RQ1.2}) How does code complexity affect the success rate of translation? We look at how lines of code and number of functions in a benchmark influence the success rate. We show lines of code strongly influences success rates.

(\textit{RQ1.3}) How idiomatic is the Rust produced by LLMs? We run Clippy~\cite{clippy}, Rust's standard linter, on the successful translations, and analyze the rates of different categories of linter warnings. We show that LLMs occasionally (1--15\% of the time) produce code with linter warnings, suggesting that the translations could be made more performant, concise, or that they use unsafe code.

\textit{\textbf{RQ2: How effective are feedback strategies at fixing translation bugs?}}
In addition to overall translation success rates, we record the initial success rates---the rate at which the first translation passes the fuzzer---and compare this to the overall success rate. We answer two concrete questions.

(\textit{RQ2.1}) How much do feedback strategies increase the translation success rate? We compare overall success rates directly to initial success rates. We show that the most effective feedback strategy increases the success rate by 7--21\% across all of our benchmarks.

(\textit{RQ2.2}) Which feedback strategies increase success rates the most? We compare the increase in success rates for each feedback strategy. We show that, surprisingly, \restart and \hinted outperform \repair and \capr consistently. We provide a plausible explanation for this result.

\textit{\textbf{RQ3: How do LLM translations compare to rule-based translation tools?}} We compare LLM translations to translations produced by the rule-based translation tool C2Rust~\cite{c2rust}. While C2Rust theoretically can guarantee the correctness of the translation, we show LLMs produce far more concise and idiomatic translations.

\textit{\textbf{RQ4: Why do translations fail?}} Translation can fail for several reasons beyond the fuzzer finding counterexamples. We report failure rates for different failure reasons.

\subsection{Experimental Setup}
\subsubsection{Implementation}
We implement \tool{} as a framework, which is instantiated with an LLM and a feedback strategy.
Once instantiated, \tool{} takes as input (1) a source program to translate and (2) a budget. 
\tool currently supports C and Go for the input program, but can also be extended to support for new source languages. 
\tool outputs either a corresponding Rust translation that passes the fuzzer, or it fails with an error. Algorithm~\ref{alg:genworacle} is used for the implementation of \tool. \tool is written entirely in Python, except for the fuzzer, which is written in Rust. 
We use GNU Parallel~\cite{tange_2024_10558745} to run experiments in parallel.

\subsubsection{LLMs}
We focus on LLMs that are the highest performing on coding tasks, which are typically proprietary LLMs hosted by third party providers. We use five LLMs in our evaluation: GPT-4o~\cite{achiam2023gpt}, Claude 2.1~\cite{claude}, Claude 3 Sonnet~\cite{claude}, Gemini Pro~\cite{gemini}, and Mixtral~\cite{jiang2024mixtral}. The first four are likely very large (1T+ parameters). On the other hand, Mixtral is comparatively small (45B parameters), but is known for performing well on coding tasks, and costs less than the others. We access GPT-4o and Gemini Pro through OpenAI's and Google's APIs. We access Claude and Mixtral through AWS Bedrock. 

\subsubsection{Benchmarks}
We collect benchmarks from real-world projects hosted on GitHub. We use the following criteria for selecting benchmarks, listed in order of priority:
(1) The project is a C or Go project. As mentioned previously, C, Go, and Rust are typically used for lower-level programming tasks, thus they are likely candidates for translation to Rust.
(2) The project only uses standard libraries. This is to limit the scope of our work. We leave translating code that uses third party libraries as future work.
(3) The project uses language features generally not present in benchmarks from prior work, such as global variables, user defined dynamically allocated data structures, pointers, error handling, etc...
(4) The project is active on GitHub. For example, it has more than 50 stars and/or 20 forks. However, we do not treat this as a hard requirement.


We choose seven projects with the aim of satisfying our criteria, and getting a diverse set of application domains. Our projects are: 
\begin{itemize}
    \item \textbf{ACH}~\cite{ach}: a Go library implementing a reader, writer, and validator for banking operations
    \item \textbf{geo}~\cite{geo}: a math-focused Go library implementing common geometry functions 
    \item \textbf{libopenaptx}~\cite{openaptx}: a C library for audio processing
    \item \textbf{opl}~\cite{opl}: a C library for sound card emulation
    \item \textbf{go-gt}~\cite{go-gt}: a Go library for graph algorithms
    \item \textbf{go-edlib}~\cite{gohyphenedlib}: a Go library string comparison and edit distance algorithms
    \item \textbf{triangolatte}~\cite{triangolatte}: a 2D triangulation library in Golang
\end{itemize}


As we will show in our experiments, LLMs are still not capable of translating entire projects, owing to their size. To create benchmarks of manageable size, we develop a tool for automatically extracting benchmarks from these projects. Our tool takes as input the entire project and a specific function identifier $f$ in the project. The tool then analyzes the project to find all of $f$'s dependencies, including all functions called by $f$ (including transitive calls), type definitions, standard libraries, global variables, etc. and extracts them intro a single, compilable file. The translation task is then to write a compilable Rust file with a function equivalent to $f$'s behavior. Our methodology for selecting benchmarks is to iterate over all functions in a project, create a benchmark for it, and keep it if it meets the following criteria: (1) it does not use 3rd party libraries, (2) the maximum depth of the call graph is less than 4.

Details on the benchmarks are given in \Cref{tab:benchmarks}. The total number of benchmarks extracted from each project is given in the column ``\#Benchs''. LoC and number of functions for individual programs vary from 13 to 597 and from 1 to 25, respectively. 

\begin{table}[]
\caption{Benchmark details}
\label{tab:benchmarks}
\centering
\resizebox{\columnwidth}{!}{\begin{tabular}{@{}rcccccc@{}}
\toprule
Project & Lang. & \#Benchs & Min\textbf{/}Max\textbf{/}Avg LoC & Min\textbf{/}Max\textbf{/}Avg \#Func & Stars & Forks \\ \midrule
\textit{libopenaptx} & C & 31 & 13 \textbf{/} 173 \textbf{/} 69 & 1 \textbf{/} 9 \textbf{/} 2.9 & 124 & 29 \\
\textit{opl} & C & 81 & 19 \textbf{/} 460 \textbf{/} 67 & 1 \textbf{/} 15 \textbf{/} 2.8 & 255 & 37 \\
\textit{go-gt} & Go & 43 & 9 \textbf{/} 213 \textbf{/} 51 & 1 \textbf{/} 16 \textbf{/} 3.5 & 10 & 2 \\
\textit{go-edlib} & Go & 36 & 13 \textbf{/} 597 \textbf{/} 62 & 1 \textbf{/} 25 \textbf{/} 3.1 & 458 & 23 \\
\textit{ach} & Go & 121 & 43 \textbf{/} 194 \textbf{/} 64 & 3 \textbf{/} 7 \textbf{/} 3.4  & 442 & 145 \\
\textit{geo} & Go & 67 & 13 \textbf{/} 70 \textbf{/} 35 & 3 \textbf{/} 7 \textbf{/} 4.1  & 1600 & 179 \\
\textit{triangolatte} & Go & 29 & 9 \textbf{/} 164 \textbf{/} 38 & 1 \textbf{/} 10 \textbf{/} 2.5 & 36 & 3 \\
\bottomrule
\end{tabular}}
\vspace{-0.3cm}
\end{table}

\subsubsection{LLM Hyperparameters} 
All LLMs use a \textit{temperature} parameter for controlling the randomness/creativity of its output. To make our results more deterministic, we use a lower temperature (i.e. less random) of 0.2. Other hyperparameters, e.g.~topP and topK, are set to the default value recommended by the LLM's provider.

\subsubsection{\tool{} Hyperparameters} 
We set the budget $b$ in \Cref{alg:genworacle} to 5. For the \hinted and \repair strategies we provide 4 examples in the prompt (more examples appeared to reduce efficiency as the context window grew). For the \capr strategy, we keep conversation window size as 3, which means that only the latest 3 incorrect translations are made available to the LLM. A translation is deemed equivalent if 5 minutes of fuzzing does not return any counterexamples.


\subsubsection{Compute Resources}
We run our experiments on a machine with an AMD EPYC 7R13 Processor with 192 cores and 380\,GB of RAM. Each translation task is run sequentially in a single thread (we do not parallelize individual translation tasks or fuzzing). As previously mentioned, all LLMs are accessed through APIs provided by third party services.

\begin{figure*}[ht]

\hspace*{\fill}%
\begin{minipage}[b]{0.4\linewidth}
\centering
\vspace{0pt}
\includegraphics[width=\textwidth]{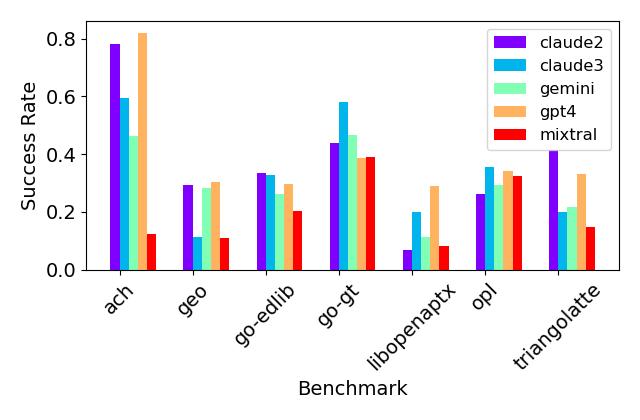}
\caption{Success rate for each LLM on each benchmark. Averaged across all feedback strategies.}
\label{fig:figure1}
\end{minipage}%
\hspace{0.5cm}
\begin{minipage}[b]{0.4\linewidth}
\centering
\vspace{0pt}
\includegraphics[width=\textwidth]{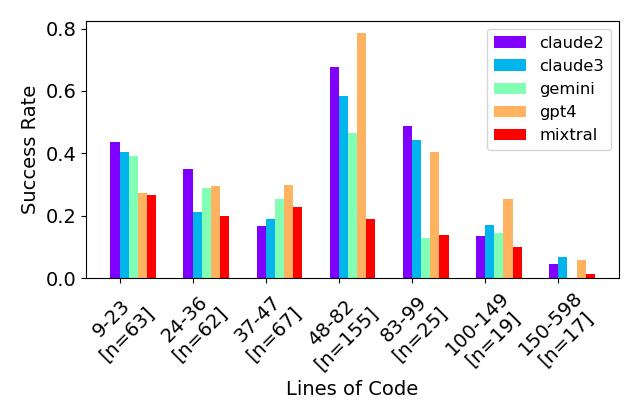}
\caption{Success rate for each LLM on benchmarks grouped by lines of code.}
\label{fig:figure2}
\end{minipage}%
\hspace*{\fill}

\end{figure*}

\begin{figure*}[ht]
\begin{minipage}[b]{0.4\linewidth}
\centering
\includegraphics[width=\textwidth]{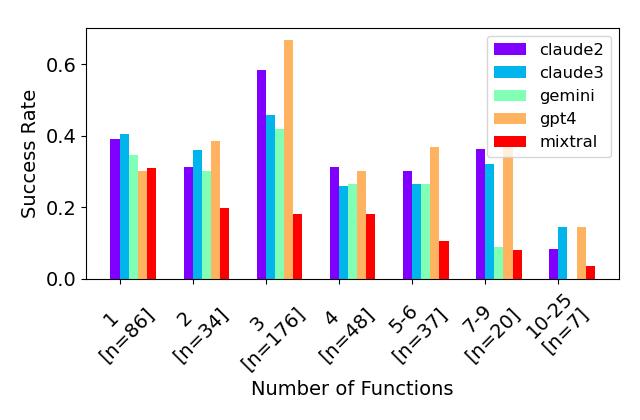}
\caption{Success rate for each LLM on benchmarks grouped by number of functions.}
\label{fig:figure3}
\end{minipage}
\hspace{0.5cm}
\begin{minipage}[b]{0.4\linewidth}
\centering
\includegraphics[width=\textwidth]{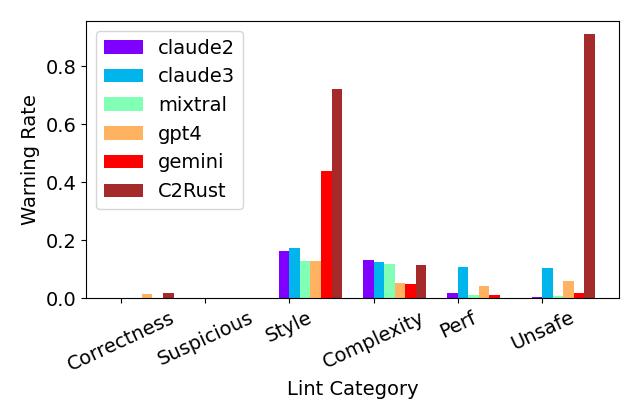}
\caption{Rates of different types of linter warnings for each LLM.}
\label{fig:figure4}
\end{minipage}
\end{figure*}

\subsection{Results}\label{sec:results}
We run a translation experiment for each of our five LLMs, four feedback strategies, and 408 benchmarks for a total of 8160 translation experiments. To account for LLM non-determinism, we run each experiment three times, and average reported metrics accross these three runs. A~translation is successful if it compiles and passes the fuzzer. A translation is failed if it: (1) does not compile, (2) the fuzzer cannot de/serialize the types used in the translation, or (3) the fuzzer finds a counterexample in the translation and the budget is reached if applicable. We answer our research questions using these results.
\\\\
\noindent\textit{\textbf{RQ1: How do LLMs perform on translating code taken from real-world projects to Rust?}}
Our LLMs achieve overall success rates of 
47.3\% (GPT-4o), 44.2\% (Claude 2), 38.5\% (Claude 3), 33.8\% (Gemini Pro), and 19.5\% (Mixtral).
We present detailed results for each LLM in Figures~\ref{fig:figure1},~\ref{fig:figure2}, ~\ref{fig:figure3}, and ~\ref{fig:figure4}. The success rate is the total number of successful translations divided by the total number of translation experiments in the category (experiments for different feedback strategies are averaged together). We answer our sub-questions below.

(\textit{RQ1.1}) How many benchmarks can each LLM translate from each of our projects? Figure~\ref{fig:figure1} shows success rates by benchmark and LLM. The best LLMs achieve success rates of 20--60\% depending on the benchmark, with one outlier of around 80\% by GPT-4o and Claude~2 on ACH. The outlier is in large part due to $\sim$40 extremely similar benchmarks in ACH, which GPT-4o and Claude 2 nearly always get right. If we remove these similar benchmarks, the success rates for GPT-4o and Claude~2 drop to around 55\%, which is in line with the other LLMs. A consistent trend is that Mixtral, while somewhat capable, has 5--20\% lower success rates than the other much larger and more expensive LLMs. However, the cost of running Mixtral (both in dollars and compute) is at least 10x less than the other LLMs. Other trends are that Claude~2, Claude~3, and GPT-4o perform similarly on most benchmarks, and they outperform Gemini in most cases.

(\textit{RQ1.2}) How does code complexity affect the success rate of translation? We use lines of code and number of functions as proxy metrics for complexity, and we show success rates for benchmarks grouped by level of complexity in Figures~\ref{fig:figure2} and~\ref{fig:figure3}. The general trend is that increasing complexity, especially in lines of code, reduces success rate. The spikes for 3 functions and 48-82 lines of code are again due to the ACH benchmarks mentioned in the previous research question. Removing these flattens the spike. In particular, success rates tend to drop off somewhere around 100+ lines of code. We discuss approaches for handling larger benchmarks in section~\ref{sec:largecode-disucssion}.

(\textit{RQ1.3}) How idiomatic is the Rust produced by LLMs? Figure~\ref{fig:figure4} shows the rate of different categories of linter warnings produced by Clippy~\cite{clippy}, Rust's standard linter. We limit our analysis to successful translations. Clippy reports five types of warnings, and we add \textbf{unsafe}. We describe them below, and give specific examples of the warnings most frequently reported by Clippy on the Rust translations.

\textbf{Correctness:} reports code that may have correctness bugs. The common examples we find are: checking if an unsigned integer is greater than 0, 
and using \lstinline[language=Rust]{MaybeUninit::uninit().assume_init()} (i.e. assuming that potentially unitialized data is initialized). 
    
\textbf{Suspicious:} the same as Correctness, but could be a false positive.
    
\textbf{Style:} code that is unidiomatic, but still correct. The common examples we find are: not following naming conventions, 
unnecessary borrows, 
using \lstinline{return} statements, 
unnecessary closure expressions (e.g. \lstinline[language=Rust]{xs.map(|x| foo(x))} instead of \lstinline[language=Rust]{xs.map(foo)}), 
using class types (e.g. \lstinline{String}) when a simple primitive type will suffice (e.g. \lstinline{str}), 
and not using idiomatic statements (e.g. using \lstinline{x <= z && z <= y}  instead of \lstinline{(x..y).contains(z)}). 
    
\textbf{Complexity:} code that could be simplified. Common examples are: 
unnecessary casting or type conversion, 
unnecessary parentheses, 
and unnecessarily putting a \lstinline[language=Rust]{Box<..>} around the type of a function parameter. 
    
\textbf{Performance:} code that could be written to run faster. The most common example is unnecessarily putting a \lstinline[language=Rust]{Box<..>} around local variables 
or collection types (e.g. \lstinline[language=Rust]{Vec}). 

\textbf{Unsafe:} code wrapped in an \lstinline[language=Rust]{unsafe} block.

Overall, LLMs produce very few correctness warnings, however they occasionally (1--15\% of the time) produce code that could be more idiomatic (Style warnings), more concise (Complexity warnings), or more performant (Performance warnings). Gemini's high rate of Style warnings is due to its preference for using \lstinline{return} statements when not necessary. We suspect that a large number of these warnings could be eliminated through prompting (e.g. instructing the LLM to prefer not using return statements, follow specific naming conventions, and not use \lstinline[language=Rust]{Box<..>} in certain cases). We also observe occasional use of \lstinline[language=Rust]{unsafe}, however the use of unsafe code was not necessary in those cases.\\

\begin{figure}
    \centering
    \includegraphics[width=0.4\textwidth]{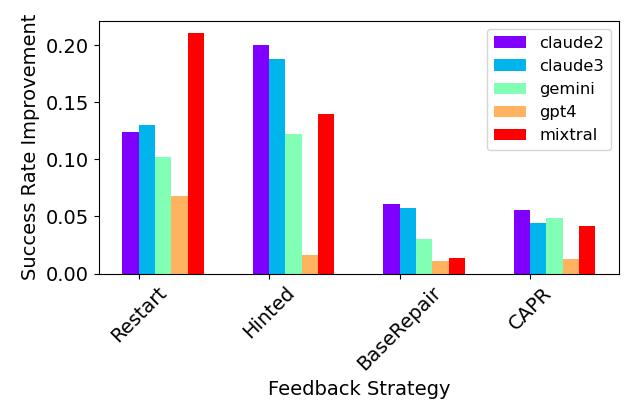}
    \vspace{-0.2cm}
    \caption{Proportional improvement in success rates after applying feedback strategies as compared to initial success rate.}
    \label{fig:figure5}
    \vspace{-0.5cm}
\end{figure}

\noindent\textit{\textbf{RQ2: How effective are feedback strategies at fixing translation bugs?}}
We answer this question by comparing the initial success rate---the rate at which the first translation passes the fuzzer, after fixing compilation errors---to the final success rate after applying feedback strategies to the unsuccessful translations. 

(\textit{RQ2.1}) How much do feedback strategies increase the translation success rate? Figure~\ref{fig:figure5} shows the improvement ratios. The best feedback strategy improves success rates by 7--21\% across all of our benchmarks. 

(\textit{RQ2.2}) Which feedback strategies increase success rates the most? Figure~\ref{fig:figure5} also shows, surprisingly, that restarting from scratch (as in \restart and \hinted) is more effective than trying to repair (as in \repair and \capr). In fact, our results suggest that providing counterexamples in the prompt may actually confuse the LLM. We discuss this trend further in Section~\ref{sec:feedback-disucssion}.
\\\\
\noindent\textit{\textbf{RQ3: How do LLM translations compare to rule-based translation tools?}} We compare the idiomatic-ness of LLM generated Rust to C2Rust~\cite{c2rust} on our opl benchmark (C2Rust failed to produce code for most of libopenaptx). Rates of linter warnings are presented in Figure~\ref{fig:figure4}. Overall we can see that the majority of code produced by C2Rust is unsafe and it is far less idiomatic, as indicated by the rate of style warnings. In addition, we observe that C2Rust produces far more verbose Rust than LLMs. On average C2Rust translations have 1.98x more LoC than LLM translations. 
\\\\
\noindent\textit{\textbf{RQ4: What is the main cause of translation failure?}} There are three reasons a translation can fail. (1) A compiling translation cannot be found. This accounts for only 7.0\% of failures. (2) The fuzzer cannot de/serialize the data types. These account for 52.6\% of failures. (3) A counterexample is found in the final translation. These account for 40.3\% of failures. The implication of this result is that we likely under-report the true translation success rate, because at least some serialization failures might be successes.

\subsection{Discussion \& Future Work}
\subsubsection{Improving Feedback Strategies}\label{sec:feedback-disucssion}
Our results suggest that prompting the LLM to repair counterexamples harms performance, which contradicts several recent works' results~\cite{PanICSE24,jana2023attention,xia2023automated,kong2024contrastrepair}. We note that two of the works~\cite{PanICSE24, jana2023attention} do not compare with a simple baseline like \restart, so they cannot conclude if this strategy helped or hurt. However, the other two~\cite{xia2023automated,kong2024contrastrepair} do report benefit from counterexamples relative to a baseline, and they evaluate on real-world code (though their task is automated program repair as opposed to code translation). Two factors likely contribute to this. First, inputs to functions in our benchmarks (and therefore the counterexamples) tend to be quite large, reflecting the complexity of the data structures used in our benchmarks (as illustrated by the examples in Section~\ref{section:intro}). Second, these inputs are generated randomly by our cross-language fuzzer. These factors result in large counterexamples (in terms of their textual representation) with random values. Our own manual analysis reveals that random fuzzer inputs are not intuitive to a human, suggesting they would not be intuitive to an LLM as well. A future direction is to study what types of counterexamples are useful for LLMs. Studying input selection and input reduction techniques would likely be immediately fruitful. 


\subsubsection{Handling Larger Benchmarks}\label{sec:largecode-disucssion}
We conjecture that the stochastic nature of LLMs' next token prediction poses fundamental limitations for translating large source programs in one go. Larger input source programs require more Rust code to be generated by the LLM, or in other words, more tokens to be predicted. Each time a token prediction is made, there is some probability that an erroneous prediction is made. Letting $e$ be the probability of an erroneous prediction, the probability that the LLM correctly predicts $n$ tokens is $(1 - e)^n$. This probability quickly goes to 0 as $n$ increases. A future direction that (at least theoretically) solves this problem is to develop a solution that partitions the input source program into chunks that can be individually translated and validated. The most obvious way to partition an input program is by function, but one could imagine even going down to the basic block level.


\section{Threats to Validity}
\label{section:threats}
Our main results are that (1) LLMs can translate real-world code, and that (2) repairing with counterexamples is not an effective feedback strategy. We discuss threats to the validity of these conclusions. The biggest threat to (1) is that the fuzzer may miss counterexamples. While we acknowledge this limitation, we point out that the fuzzer achieves 97\% coverage on average, thus we are confident that the translations are ``mostly'' correct. This limitation is also generally accepted in prior work, which uses test suites to assess correctness. Another threat to (1) is that our results do not generalize to other languages. We argue this is unlikely for popular languages like Java and Python, given that they more represented in LLM training data than C and Go. However, even if our results do not generalize to other languages, translating C and Go to Rust is still highly practical. The biggest threat to (2) is that other prompting strategies may improve the LLMs ability to use counterexamples. While possible, we believe this is unlikely due to the complexity of inputs in our benchmarks, as explained in Section~\ref{sec:feedback-disucssion}. Finally, recent work~\cite{ouyang2023llm} has shown high non-determinism in code generated by LLMs, which poses a threat to both (1) and (2). We address this by running our experiments 3 times and averaging across runs. 
\section{Conclusion}
\label{section:conclusion}
In this work, we study the ability of LLMs to translate real-world code to Rust. We present \tool{}, an end-to-end Rust transpiler, and we use it to test the ability of five state-of-the-art LLMs to translate C and Go code taken from real-world projects to Rust. Our results demonstrate that LLMs are indeed capable of translating code to Rust, though there is room for improvement. 
In addition, we show that counterexamples, at least random fuzzer generated counterexamples, are ineffective feedback for an LLM.

\bibliographystyle{ACM-Reference-Format}
\bibliography{reference}

\end{document}